# The "forgotten" pseudomomenta and gauge changes in generalized Landau Level problems: spatially nonuniform magnetic and temporally varying electric fields

*Georgios Konstantinou[1] and Konstantinos Moulopoulos[2]*

*Department of Physics, University of Cyprus, PO Box 20537, 1678 Nicosia, Cyprus*

[1]ph06kg1@ucy.ac.cy, [2]cos@ucy.ac.cy

By perceiving gauge invariance as an analytical tool in order to get insight into the states of the "generalized Landau problem" (a charged quantum particle moving inside a magnetic, and possibly electric field), and motivated by an early article that correctly warns against a naive use of gauge transformation procedures in the usual Landau problem (i.e. with the magnetic field being static and uniform), we first show how to bypass the complications pointed out in that article by solving the problem in full generality through gauge transformation techniques in a more appropriate manner. Our solution provides in simple and closed analytical forms all Landau Level-wavefunctions without the need to specify a particular vector potential. This we do by proper handling of the so-called pseudomomentum $\vec{K}$ (or of a quantity that we term pseudo-angular momentum $L_z$), a method that is crucially different from the old warning argument, but also from standard treatments in textbooks and in research literature (where the usual Landau-wavefunctions are employed - labeled with canonical momenta quantum numbers). Most importantly, we go further by showing that a similar procedure can be followed in the more difficult case of spatially-nonuniform magnetic fields**:** in such case we define $\vec{K}$ and $L_z$ as plausible generalizations of the previous ordinary case, namely as appropriate line integrals of the inhomogeneous magnetic field – our method providing closed analytical expressions for all stationary state wavefunctions in an easy manner and in a broad set of geometries and gauges. It can thus be viewed as complementary to the few existing works on inhomogeneous magnetic fields, that have so far mostly focused on determining the energy eigenvalues rather than the corresponding eigenkets (on which they have claimed that, even in the simplest cases, it is not possible to obtain in closed form the associated wavefunctions). The analytical forms derived here for these wavefunctions enable us to also provide explicit Berry's phase calculations and a quick study of their connection to probability currents and to some recent interesting issues in elementary Quantum Mechanics and Condensed Matter Physics. As an added feature, we also show how the possible presence of an additional electric field can be treated through a further generalization of pseudomomenta and their proper handling.

## *1. Introduction*

The enormous rising of importance in the past four decades of the Landau problem (electron inside a spatially-uniform magnetic field) – as the key problem in the general area of the Quantum Hall Effect but also as highly relevant to the recent burst in topological phases – and the central role of gauge invariance in all of Physics (to be recalled below), can make one wonder whether the latter can be used methodologically in order to provide insight into the solutions of the former. In this respect, and by additionally being motivated by an early article [1] that points out very justifiable warnings against a naive handling of gauge transformation procedures in the above Landau problem, especially when passing from one Landau gauge to another – and also by realizing the almost exclusive use of the original method of Landau (namely, the use of canonical momentum operators to provide good quantum numbers) in textbooks but also in research literature -- we present here an alternative procedure, not based on canonical momenta, by placing emphasis on the proper use of (the generator of translations inside a magnetic field) pseudomomentum $\vec{K}$ in Cartesian coordinates (and of a rotational type of quantity $L_z$, when working with polar coordinates, that we term pseudo-angular momentum). Our method completely eliminates the complications pointed out in the old article (as well as the risk of errors in all standard treatments that involve the usual Landau-wavefunctions) and provides in simple and closed analytical forms all stationary wavefunctions in any arbitrary gauge. Through our solution it is indeed demonstrated that the book-keeping provided by the use of the pseudomomenta is superior (to canonical momenta) especially in cases of energy degeneracy (as occurs here with the Landau Levels), in that there is no necessity of changing the basis associated with changing to different components of pseudomomentum that label differently the degenerate states, as actually happens with the ordinary Landau problem and had been pointed out in [1].

In the present work we go much further than the ordinary uniform field case mentioned above, by also making similar (but extended) considerations on the "generalized Landau problem" with a nonuniform magnetic field (but also with an additional electric field, as we shall see). However, let us first recall the issue of gauge invariance, in its full generality and in a "practical" manner (i.e. in a way that is usable to provide insight into the solutions of all these problems)**:** It is well-known from Weyl's early work [2] (and also from independent proposals by Schrodinger (1922), Fock (1927) and London (1927) [3]), that the structure of the time-dependent Schrodinger equation (TDSE) remains the same upon change of potentials (through $\vec{A}_2 = \vec{A}_1 + \vec{\nabla}\Lambda$ and $V_2 = V_1 - \frac{1}{c}\frac{\partial \Lambda}{\partial t}$) if at the same time the wavefunction $\Psi_1(\mathbf{r},t)$ (solution of the TDSE for the set of potentials ($\vec{A}_1$, $V_1$)) is replaced by $\Psi_2(\mathbf{r},t) = \Psi_1(\mathbf{r},t)e^{-i\frac{e\Lambda}{\hbar c}}$. Hence, the *formal* solution (meaning**:** before any imposition of boundary conditions) of the TDSE for the set of potentials ($\vec{A}_2$, $V_2$) is the above $\Psi_2$. This formal connection between two systems (in which the same particle of charge -*e* moves in the above two different sets of potentials) may be seen as a *mapping* between the two problems, and this has been exploited recently [5,6] for advancing new solutions of *t*-dependent Aharonov-Bohm configurations, both of the magnetic and the electric type, an area that after [5,6] seems to be growing rapidly [7]. In a similar manner, if we look at the *t*-independent Schrodinger equation (TISE), we can see that a similar formal mapping is also valid for the stationary state solutions, i.e. $\Psi_2(\mathbf{r}) = \Psi_1(\mathbf{r})e^{-i\frac{e\Lambda}{\hbar c}}$ (now only under the change $\vec{A}_2 = \vec{A}_1 + \vec{\nabla}\Lambda$, with $\Lambda$ now being time-independent), with the understanding that the corresponding solutions $\Psi_1$ and $\Psi_2$ will belong *to the same energy*. Long time ago Swenson [1] criticized the above result, pointing out that there may be subtleties associated with cases of *energy degeneracies*; and in particular for the Landau problem, he demonstrated that it is incorrect to apply the above simple result to the well-known wavefunction-solutions of the TISE in the two standard Landau gauges ($\vec{A}_1 = -yB\hat{i}$ and $\vec{A}_2 = xB\hat{j}$, with *B* the modulus of the uniform magnetic field (with direction along the positive *z*-axis) and $\hat{i}, \hat{j}$ unit vectors along *x,y* directions). He actually showed that the mapping between the two problems is more complicated (involving superposition of different $\Psi_1$'s that corresponded to different values of a canonical momentum but the same energy as the one of $\Psi_2$), and he then drew the general conclusion that in cases of degeneracy the simple mapping (with the above single extra phase factor) produces a solution that is not necessarily a single stationary state of the 2$^{nd}$ system.

Here we want to point out that if one is careful to map states that correspond not only to the same energy but simultaneously to the same value of another constant of motion, such as a *given component* of pseudomomentum (which is chosen to be *the same component* in both gauges), then the simple mapping is valid again. In Swenson's case (and in almost all textbooks as well as in most of the research literature), it so happens that the $\Psi_1$ and $\Psi_2$ used (the ones based on canonical momenta) that are the standard solutions of this problem in the usual two Landau gauges, *do not* correspond to the same Cartesian component of pseudomomentum (as will be recalled below), hence the arising of a more complicated result. The use of the same component of another constant of motion indeed simplifies the problem, and this simplification can be exploited, as we will see, in order to immediately write down analytical solutions for any gauge, and finally to go further by drawing general conclusions on the problem of spatially non-uniform magnetic fields where analytical results are rare. In this case we find it advantageous to introduce a plausible generalization of the pseudomomenta and pseudo-angular momenta, and with their proper use we provide closed analytical forms of all stationary state wavefunctions for a particular example of an inhomogeneous magnetic field. Determination of the corresponding energy eigenvalues is not attempted here, but we do provide the rather few references that do this through transfer matrix techniques – our work been complementary to the others in the sense that it provides immediate information on the eigenkets, that are claimed to be difficult (or even impossible) to obtain by workers in this field. We even take advantage of the closed analytical forms of the wavefunctions in order to study certain Berry's phase effects upon an adiabatic and cyclic movement of the origin, and to connect the results to probability currents and to several open problems associated with probability flow in the theory of elementary Quantum Mechanics and Solid State Physics. Finally, we demonstrate how the method can be generalized in the additional presence of an electric field, which we show that can even be temporally-varying without major changes. In this article we everywhere assume for simplicity a spatially 2-D system (in Cartesian (*x,y*) but also in polar (*r,φ*) coordinates), with the magnetic field (either uniform or non-uniform) always being perpendicular to the 2-D system (i.e. along the *z* direction) and with the electric field, whenever present, lying in the 2-D (*x,y*) plane. Also for simplicity we everywhere choose static vector potentials ($\partial \vec{A}/\partial t = 0$).

## 2. Homogeneous magnetic field

### 2.1 Cartesian coordinates

A central quantity in the Landau problem (for a quantum particle of charge $-e$), although a bit "forgotten" through the years (in favor of the use of guiding center operators in this problem – see later below), is what had been earlier called the pseudomomentum [8], defined as

$$\vec{K} = \vec{p} + \frac{e}{c}\vec{A} + \frac{e}{c}\vec{r} \times \vec{B} \quad \textbf{(2.1)}$$

where $\vec{B} = B\hat{z}$ and $\vec{A}$ is the vector potential defined by $\vec{\nabla} \times \vec{A} = \vec{B}$ **(2.2)**, which apart from (2.2) we intentionally leave unspecified (but static). For a classical system it is immediately obvious that the vector quantity $\vec{K}$ is indeed a constant of motion: from Newton's equation of motion with the Lorentz force, namely $d\vec{\Pi}/dt = d(\vec{p} + e\vec{A}/c)/dt = -e\vec{v} \times \vec{B}/c$, by simply writing $\vec{v} = d\vec{r}/dt$ we immediately obtain $d\vec{K}/dt = 0$. For a quantum system like the one we are interested in, it is also straightforward to see that this vector operator quantity $\vec{K}$ is a constant of motion, namely $[H, \vec{K}] = 0$ (and with a little more effort one can show that it is also the generator of translations inside a uniform magnetic field). [As already mentioned, this pseudomomentum is rarely mentioned nowadays in the area of the Quantum Hall Effect, having yielded its place to the so-called guiding center operators $(X_0, Y_0)$, or $\vec{R}_0 = X_0\,\hat{i} + Y_0\,\hat{j}$, the relation between them being $\vec{K} = -e\vec{B} \times \vec{R}_0/c$ – see later below.] Our target is to find common eigenstates between $H$ and the pseudomomentum $\vec{K}$ for any vector potential that satisfies (2.2) (something that, although it sounds natural, is never actually followed, the standard procedure being the one of Landau [9], with the use of *canonical* momentum, which is *not* a vector constant of the motion – only one of its Cartesian components being conserved, and this being a *different* component in different gauges (see below)). Basing therefore our work on the pseudomomentum as a better book-keeping of the constants of motion, let us first choose the *x*-component of pseudomomentum $K_x = p_x + \frac{e}{c}A_x + \frac{e}{c}yB$ and solve its eigenvalue equation, namely

$$K_x\Psi(x,y) = k_x\Psi(x,y) \Rightarrow -i\hbar\frac{\partial\Psi}{\partial x} + \frac{e}{c}yB\Psi + \frac{e}{c}A_x(x,y)\Psi = k_x\Psi \quad \textbf{(2.3)}$$

where $k_x$ is the continuous eigenvalue of $K_x$. The solution of the above differential equation is

$$\Psi(x,y) = Ce^{i\left(\frac{k_x}{\hbar} - \frac{eyB}{\hbar c}\right)x} e^{-\frac{ie}{\hbar c}\int_0^x dx'\,A_x(x',y)} f(y) = Ce^{i\frac{k_x x}{\hbar}} e^{-i\frac{eyBx}{\hbar c}} e^{-\frac{ie}{\hbar c}\int_0^x dx'\,A_x(x',y)} f(y) \quad \textbf{(2.4)}$$

with $f(y)$ a function of $y$, to be determined. This wavefunction must also satisfy the static Schrodinger equation $H\Psi = E\Psi$ with Hamiltonian that contains a minimal substitution (namely the kinematic momentum $\Pi = p + \frac{e}{c}A$ in place of the canonical momentum $p$), which upon expansion of $\Pi^2$ gives

$$H = \frac{p^2}{2m} + \frac{e}{mc}\vec{A}\cdot\vec{p} - i\hbar\frac{e}{2mc}\vec{\nabla}\cdot\vec{A} + \frac{e^2}{2mc^2}A^2$$

for an arbitrary vector potential $\vec{A}$ that satisfies eq. (2.2). This leads, after several algebraic manipulations, to the equation that $f(y)$ must satisfy, namely

$$-\frac{\hbar^2}{2m}\frac{\partial^2 f}{\partial y^2} - iA_y(0,y)\frac{\hbar e}{mc}\frac{\partial f}{\partial y} + \left[\frac{1}{2m}\left(k_x - \frac{eyB}{c}\right)^2 + \frac{1}{2m}\left(\frac{e}{c}\right)^2 A_y^2(0,y) - i\frac{\hbar e}{2mc}\frac{\partial A_y(0,y)}{\partial y}\right]f = Ef \ .$$

**(2.5)**

Finally, upon change of variable according to $f(y) = e^{-\frac{ie}{\hbar c}\int_0^y dy' A_y(0,y')} Y(y)$ all terms containing components of vector potential cancel out and we obtain a very simple equation that $Y(y)$ must satisfy, namely

$$-\frac{\hbar^2}{2m}\frac{\partial^2 Y}{\partial y^2} + \frac{1}{2m}\left(k_x - \frac{eyB}{c}\right)^2 Y = EY \ , \textbf{(2.6)}$$

which is a one-dimensional shifted harmonic oscillator with the well-known solutions, Hermite polynomials times a Gaussian, and with the standard harmonic oscillator energies (but infinitely degenerate for an infinite system), namely $\varepsilon_n = \hbar\omega(n+1/2)$ (with $n$ the non-negative Landau Level index), and with $\omega = \frac{eB}{mc}$ the cyclotron frequency). The total solution can then be written (from (2.4)) as

$$\Psi(x,y) = Ce^{i\left(\frac{k_x}{\hbar} - \frac{eyB}{\hbar c}\right)x} e^{-\frac{ie}{\hbar c}\int_0^x dx' A_x(x',y)} f(y) = Ce^{i\frac{k_x x}{\hbar}} e^{-i\frac{eyBx}{\hbar c}} e^{-\frac{ie}{\hbar c}\left[\int_0^x dx' A_x(x',y) + \int_0^y dy' A_y(0,y')\right]} Y(y) \ ,$$

$$= e^{-i\frac{e\Lambda}{\hbar c}} e^{i\frac{k_x x}{\hbar}} Y(y)$$

**(2.7)**

with $Y(y) \sim H_n\left(\frac{y-Y_o}{l_B}\right) e^{-\frac{(y-Y_o)^2}{2l_B^2}}$ and $Y_o = ck_x / eB = k_x l_B^2$ being the well-known guiding center operator eigenvalue (and with $l_B$ denoting the so-called magnetic length). [We remind the reader that, in general, the guiding center vector operator $\vec{R}_0 = (X_0, Y_0)$ (defined through the same combination of positions and momenta that give the point $(X_0, Y_0)$ of the center of the classical circle, namely $X_0 = x + \Pi_y/m\omega$ and $Y_0 = y - \Pi_x/m\omega$, with $\Pi_x, \Pi_y$ kinematic momenta and $\omega$ the cyclotron frequency $\omega = \frac{eB}{mc}$) is related to the pseudomomentum $\vec{K}$ through $\vec{K} = -e\vec{B} \times \vec{R}_0 / c$, hence $K_x = eBY_0/c$ and $K_y = -eBX_0/c$.] It is directly seen that the phase factor $\Lambda$ appearing in (2.7) contains line integrals of the vector potential (along a connected path) and a flux of magnetic field $\vec{B} = B\vec{z}$, namely

$$\Lambda(x,y) = yBx + \int_0^x dx' A_x(x',y) + \int_0^y dy' A_y(0,y') = \int_0^x \int_0^y dx dy B + \int_0^x dx' A_x(x',y) + \int_0^y dy' A_y(0,y'), \quad (2.8)$$

and one notes that its gradient is $\vec{\nabla}\Lambda(x,y) = By\hat{i} + \vec{A}(x,y)$. [Note that, although it is easy to see that indeed $\frac{\partial\Lambda}{\partial x} = yB + A_x(x,y)$, working out $\frac{\partial\Lambda}{\partial y}$ requires some more steps: $\frac{\partial\Lambda}{\partial y} = xB + \int_0^x dx' \frac{\partial A_x(x',y)}{\partial y} + A_y(0.y)$, which upon substitution $\frac{\partial A_x(x',y)}{\partial y} = \frac{\partial A_y(x',y)}{\partial x'} - B$, makes the integral turn out to be $A_y(x,y) - A_y(0,y) - xB$, which finally yields $\frac{\partial\Lambda}{\partial y} = A_y(x,y)$.] It is therefore clearly seen that (2.7) is the solution of Schrodinger equation that maps the Landau gauge $\vec{A} = -By\hat{i}$ (whose solution is the 2$^{nd}$ and 3$^{rd}$ factor of the last part of (2.7)) to an arbitrary vector potential $\vec{A}(x,y)$ in the sense discussed in the Introduction (indeed there is a gauge transformation that connects the two gauges, namely $\vec{A}(x,y) = -By\hat{i} + \vec{\nabla}\Lambda$ as already pointed out above). Note also that the form (2.8) involves a path (in the line integral of $\vec{A}$) that connects the point (0,0) to the point (x,y) (in place of (0,0) we could have more generally had an arbitrary $(x_0,y_0)$ – see later below), and apart from this line integral it also involves an additional flux contribution that has the general form of the non-local terms-results in [5,6].

Similarly, if we had chosen to find the simultaneous eigenfunctions of $H$ and $K_y$ (although note, this we will *not* need here, except possibly for comparison with the extended results of section 5 with the additional presence of an electric field), we would have ended up with

$$\Psi(x,y) = C e^{i\frac{k_y y}{\hbar}} e^{-i\frac{e}{\hbar c}\left[-\int_0^x \int_0^y dx dy B + \int_0^x dx' A_x(x',0) + \int_0^y dy' A_y(x,y')\right]} X(x),$$

with the phase factor $\Lambda(x,y) = -\int_0^x \int_0^y dx dy B + \int_0^x dx' A_x(x',0) + \int_0^y dy' A_y(x,y')$ satisfying $\vec{\nabla}\Lambda(x,y) = -Bx\hat{i} + \vec{A}(x,y)$, hence this solution mapping the second Landau gauge $\vec{A} = Bx\hat{j}$ to an arbitrary vector potential $\vec{A}(x,y)$, as expected.

So far all the results are for an arbitrary vector potential $\vec{A}$. If we now choose a specific gauge for $\vec{A}$, let us say one of the Landau gauges above, namely $\vec{A_1} = xB\hat{j}$, the solution (2.7) (which, recall, came out of simultaneous diagonalization of H and $K_x$) reads

$$\Psi^{\vec{A_1}}(x,y) = C e^{i\left(\frac{k_x}{\hbar} - \frac{eyB}{\hbar c}\right)x} e^{-\frac{ie}{\hbar c}\int_0^x dx' A_x(x',y)} f(y) = C e^{i\frac{k_x x}{\hbar}} e^{-i\frac{eyBx}{\hbar c}} Y(y) \quad (2.9)$$

And if we choose another Landau gauge, i.e. $\vec{A_2} = -yB\hat{i}$ equation (2.7) (so that we again have $K_x$ simultaneously diagonalized with H) gives

$$\Psi^{\vec{A_2}}(x,y) = C e^{i\left(\frac{k_x}{\hbar} - \frac{eyB}{\hbar c}\right)x} e^{-\frac{ie}{\hbar c}\int_0^x dx' A_x(x',y)} f(y) = C e^{i\frac{k_x x}{\hbar}} e^{-i\frac{eyBx}{\hbar c}} e^{+\frac{ieB}{\hbar c}\left[\int_0^x dx'y\right]} Y(y) = C e^{i\frac{k_x x}{\hbar}} Y(y)$$
**(2.10)**

which, incidentally, is the standard solution (in textbooks and literature) for this latter gauge.

Note then that (2.9) and (2.10) only differ by a certain phase factor, namely $\Psi^{\vec{A_2}}(x,y) = e^{i\frac{eyBx}{\hbar c}} \Psi^{\vec{A_1}}(x,y)$ **(2.11)** (this phase containing the correct $\Lambda$ for a gauge change from $\vec{A_1}$ to $\vec{A_2}$ as can be easily verified), resulting in the observation that different wavefunctions for different gauges may indeed differ by the standard single phase factor as long as the *same* component (of the constant of motion) is chosen to be diagonalized with the Hamiltonian (here the choice was $K_x$). This contradicts Swenson's [1] results on gauge changes, where he states that a wavefunction that belongs to a specific gauge must be written as a linear combination of wavefunctions in another gauge. Swenson's result is indeed true in case that in the two gauges, two *different* operators (here different components of the constant of motion) are used to be simultaneously diagonalized with the Hamiltonian (and this is exactly what happens with the standard (a la Landau [9]) use of *canonical* momenta). Here is why - and, incidentally, this is a word about readers worrying about the "other" Cartesian component of K, $K_y$ (or of the guiding center operator $X_0$)**:** in Landau gauge $\vec{A_1}$, it turns out from the definition (2.1) that simply $K_x = p_x$, and in Landau gauge $\vec{A_2}$, it turns out that $K_y = p_y$. Hence, indeed, the insistence on the use of *p*'s by Landau ($p_x$ in the 1st gauge and $p_y$ in the 2nd gauge) to simultaneously diagonalize them with H is not quite the most natural choice**;** it corresponds to *different components of pseudomomentum* being diagonalized in each of the two gauges, hence the Swenson's complications.

## 2.2 Polar coordinates

As we will see here, similar conclusions can be obtained in a slightly more nontrivial example, namely the same system in polar coordinates, where the conserved quantity turns out to be what could be called a pseudo-angular momentum $L_z = (\vec{r} \times \vec{\Pi})_z - \frac{eB}{2c} r^2$ **(2.12)** (noted in [10]), with $\vec{\Pi}$ the kinematic momentum in two dimensions. Since $[H, L_z] = 0$, this quantity remains a constant of motion for any gauge and we may choose it to be simultaneously diagonalised with $H$. Note again that this pseudo-angular momentum is defined with an arbitrary vector potential and it is gauge invariant [10]. It can be shown, after considerable algebra, that the correct – common for $H$ and $L_z$ – wavefunction may be written as

$$\Psi(r,\varphi) = C e^{i\frac{\lambda\varphi}{\hbar}} e^{i\frac{eBr^2\varphi}{2\hbar c}} e^{-\frac{ie}{\hbar c}\left[\int_0^\varphi r d\varphi' A_\varphi(r,\varphi') + \int_0^r dr' A_r(r,0)\right]} R(r) \quad (2.13)$$

where $\lambda$ is the eigenvalue of the pseudo-angular momentum operator and $R(r)$ a function of $r$ that satisfies the following differential equation:

$$-\frac{\hbar^2}{2m}\frac{1}{r}\frac{\partial}{\partial r}\left(r\frac{\partial R}{\partial r}\right) + \frac{1}{2mr^2}\left(\lambda + \frac{eBr^2}{2c}\right)^2 R = ER \quad (2.14)$$

This is the Laguerre equation expressed in polar coordinates. Note that the following commutation relation between the pseudomomentum and pseudo-angular momentum holds

$$\left[\vec{K}, L_z\right] = i\hbar\left(\hat{z} \times \vec{K}\right) \quad (2.15)$$

[This can be seen by examining the commutation of i.e. the $x$-component of pseudomomentum with the pseudo-angular momentum, namely

$$[K_x, L_z] = \left[p_x + eA_x/c + eyB/c, x\Pi_y - y\Pi_x - eBr^2/2c\right] =$$
$$= -i\hbar\left[p_y - \frac{ex}{c}\frac{\partial A_y}{\partial x} + eA_y/c + \frac{ex}{c}\frac{\partial A_x}{\partial y}\right] = -i\hbar\left[p_y + eA_y/c - \frac{exB}{c}\right] = -i\hbar K_y .]$$

Using then again a Landau gauge (although not the most natural choice in this polar description, but we still make this choice in order to demonstrate a point later below), namely $\vec{A}_1 = xB\hat{j} \Rightarrow A_\varphi = rB\cos^2\varphi$ and $A_r = rB\cos\varphi\sin\varphi$ the solution (2.13) gives

$$\Psi^{\vec{A}_1}(r,\varphi) = Ce^{i\frac{\lambda\varphi}{\hbar}} e^{-i\frac{eBr^2 \sin 2\varphi}{4\hbar c}} R(r) \quad (2.16)$$

While, in the other (perpendicular) Landau gauge $\vec{A}_2 = -yB\hat{i} \Rightarrow A_\varphi = rB\cos^2\varphi A_\varphi = rB\sin^2\varphi$, $A_r = -rB\cos\varphi\sin\varphi$, eq. (2.13) gives

$$\Psi^{\vec{A}_2}(r,\varphi) = Ce^{i\frac{\lambda\varphi}{\hbar}} e^{i\frac{eBr^2 \sin 2\varphi}{4\hbar c}} R(r). \quad (2.17)$$

We can clearly see again that (2.16) and (2.17) are again connected to each other through a simple gauge transformation that conserves the pseudo-angular momentum, namely, it is indeed true that $\Psi^{\vec{A}_2} = \Psi^{\vec{A}_1} e^{-i\frac{e\Lambda}{\hbar c}}$ with the correct gauge function $\Lambda = -Bxy = -\frac{Br^2}{2}\sin 2\varphi$ (that indeed takes us from $\vec{A}_1$ to $\vec{A}_2$).

It is also easy to see that the same method applied for the mapping from one of the Landau gauges to the symmetric gauge $\vec{A}_3 = \frac{1}{2}\vec{B} \times \vec{r}$ (which in Cartesian coordinates is $\vec{A}_3 = -yB\hat{i}/2 + xB\hat{j}/2$) will immediately give the correct $\Lambda$ that connects the two gauges, which gives a proper answer to the question posed at the end of Swenson's article, something that we leave for the reader to check.

### 3. Inhomogeneous magnetic field

In classical mechanics the motion in a inhomogeneous magnetic field $\vec{B}(\vec{r})$ is described by Newton's second law:

$$m\frac{d\vec{v}}{dt} = -\frac{e}{c}\vec{v} \times \vec{B}(\vec{r}), \quad \textbf{(3.1)}$$

where $m$ is the mass, $\vec{v}$ is the velocity and $-e$ the charge of the particle. $\vec{B}(\vec{r})$ is an $\vec{r}$-dependent static magnetic field. The energy of the particle is a conserved quantity, irrespective of the spatial structure of the magnetic field, namely

$$E = \frac{1}{2}mv^2 \Rightarrow \frac{dE}{dt} = m\vec{v} \cdot \frac{d\vec{v}}{dt} = -\frac{e}{c}\vec{v} \cdot (\vec{v} \times \vec{B}(\vec{r})) = 0, \quad \textbf{(3.2)}$$

because the particle accelerates in a direction that is always perpendicular to its velocity. However, not only the energy is conserved; under certain circumstances (i.e. with respect to the field's symmetry) there are other quantities that are also constants of the motion, like a generalized type of pseudo-angular momentum, which we now define by the relation

$$L_z = (\vec{r} \times \vec{\Pi})_z - \frac{e}{c}\int_C B(\vec{r})(\vec{r} \cdot d\vec{r}) \quad \textbf{(3.3)}$$

And naturally the pseudomomentum itself, now defined through

$$\vec{K} = \vec{\Pi} + \frac{e}{c}\int d\vec{r} \times \vec{B}(\vec{r}), \quad \textbf{(3.4)}$$

all these quantities defined in a 2D plane perpendicular to $\vec{B}(\vec{r})$ which is always oriented parallel to the *z* axis. In general, the previous constants are expressed as line integrals and therefore they cannot quite represent a well-defined constant of motion, at least in the ordinary sense. But if the integrand is a curl-free quantity, i.e. when it happens that

$$\vec{\nabla}\times\left[B(\vec{r})\vec{r}\right] = B(\vec{r})\underbrace{\vec{\nabla}\times\vec{r}}_{=0} - \vec{r}\times\vec{\nabla}B(\vec{r}) = -\vec{r}\times\vec{\nabla}B(\vec{r}) = \left(-x\frac{\partial B}{\partial y} + y\frac{\partial B}{\partial x}\right)\hat{z},\text{ or, in polar coordinates}$$

$$\vec{\nabla}\times\left[\vec{r}B(\vec{r})\right] = \frac{\partial B}{\partial \phi}\hat{z}$$

becomes zero (i.e. if the field has cylindrical symmetry) then $L_z$ becomes a well-defined constant of motion (independent of the path *C*), that (by choosing the path *C* being a straight line in the radial direction) can be written as

$$L_z = \left(\vec{r}\times\vec{\Pi}\right)_z - \frac{e}{c}\int_0^r B(r')r'dr' \quad \textbf{(3.5)}$$

The above can actually be proven directly using the commutator

$$\left[H, L_z\right] = \left[H, x\Pi_y - y\Pi_x\right] - \frac{e}{c}\left[H, \int_0^r B(r')r'dr'\right] = 0,$$

with $H = \frac{\Pi_x^2}{2m} + \frac{\Pi_y^2}{2m}$. This allows us to search for common eigenstates for $H, L_z$. Using then eq. (3.5) expressed in polar coordinates we have

$$L_z = -i\hbar\frac{\partial}{\partial \varphi} + \frac{e}{c}rA_\varphi - \frac{e}{c}\int_0^r B(r')r'dr' \quad \textbf{(3.6)}$$

and with the ansatz $\Psi(r,\varphi) = Ce^{i\frac{\lambda\varphi}{\hbar}} e^{\frac{ie}{\hbar c}\int_0^r r'dr'B(r')\varphi} e^{-\frac{ie}{\hbar c}\int_0^\varphi rd\varphi'A_\varphi(r,\varphi')} f(r)$ we can solve the eigenvalue equation $L_z\Psi(r,\varphi) = \lambda\Psi(r,\varphi)$, with $\lambda$ being an eigenvalue of $L_z$. Substituting this ansatz in the static Schrodinger equation with Hamiltonian

$$H = \frac{p^2}{2m} + \frac{e}{mc}\vec{A}\cdot\vec{p} - i\hbar\frac{e}{2mc}\vec{\nabla}\cdot\vec{A} + \frac{e^2}{2mc^2}A^2$$

$$= -\frac{\hbar^2}{2m}\nabla^2 - \frac{i\hbar e\vec{A}\cdot\vec{\nabla}}{mc} - i\hbar\frac{e}{2mc}\vec{\nabla}\cdot\vec{A} + \frac{e^2}{2mc^2}A^2$$

and with $\vec{\nabla} = \frac{\partial}{\partial r}\hat{r} + \frac{1}{r}\frac{\partial}{\partial \varphi}\hat{\varphi}$ and $\nabla^2 = \frac{1}{r}\frac{\partial}{\partial r}\left(r\frac{\partial}{\partial r}\right) + \frac{1}{r^2}\frac{\partial^2}{\partial \varphi^2} = \frac{1}{r}\frac{\partial}{\partial r} + \frac{\partial^2}{\partial r^2} + \frac{1}{r^2}\frac{\partial^2}{\partial \varphi^2}$

we get

$$\frac{\partial \Psi}{\partial r} = \frac{ie}{\hbar c} rB(r)\varphi\Psi - \frac{ie}{\hbar c}\int_0^\varphi d\varphi' \frac{\partial}{\partial r}\left[rA_\varphi(r,\varphi')\right]\Psi + Ce^{i\frac{\lambda\varphi}{\hbar}} e^{\frac{ie}{\hbar c}\int_0^r r'dr'B(r')\varphi} e^{-\frac{ie}{\hbar c}\int_0^\varphi rd\varphi' A_\varphi(r,\varphi')} \frac{\partial f}{\partial r} \quad (3.7)$$

and

$$\frac{\partial^2 \Psi}{\partial r^2} = \frac{ie}{\hbar c}\frac{\partial}{\partial r}(rB(r))\varphi\Psi + \frac{ie}{\hbar c} rB(r)\varphi\frac{\partial \Psi}{\partial r} - \frac{ie}{\hbar c}\int_0^\varphi d\varphi' \frac{\partial^2}{\partial r^2}\left[rA_\varphi(r,\varphi')\right]\Psi - \frac{ie}{\hbar c}\int_0^\varphi d\varphi' \frac{\partial}{\partial r}\left[rA_\varphi(r,\varphi')\right]\frac{\partial \Psi}{\partial r}$$

$$+ C\frac{ie}{\hbar c} rB(r)\varphi e^{i\frac{\lambda\varphi}{\hbar}} e^{\frac{ie}{\hbar c}\int_0^r r'dr'B(r')\varphi} e^{-\frac{ie}{\hbar c}\int_0^\varphi rd\varphi' A_\varphi(r,\varphi')} \frac{\partial f}{\partial r}$$

$$- C\frac{ie}{\hbar c}\int_0^\varphi d\varphi' \frac{\partial}{\partial r}\left[rA_\varphi(r,\varphi')\right]e^{i\frac{\lambda\varphi}{\hbar}} e^{\frac{ie}{\hbar c}\int_0^r r'dr'B(r')\varphi} e^{-\frac{ie}{\hbar c}\int_0^\varphi rd\varphi' A_\varphi(r,\varphi')} \frac{\partial f}{\partial r}$$

$$+ Ce^{i\frac{\lambda\varphi}{\hbar}} e^{\frac{ie}{\hbar c}\int_0^r r'dr'B(r')\varphi} e^{-\frac{ie}{\hbar c}\int_0^\varphi rd\varphi' A_\varphi(r,\varphi')} \frac{\partial^2 f}{\partial r^2}$$

**(3.8)**

Using then $B(r) = \frac{1}{r}\left[\frac{\partial}{\partial r}[rA_\varphi(r,\varphi')] - \frac{\partial}{\partial \varphi'} A_r(r,\varphi')\right]$, we find

$$rB(r) + \frac{\partial}{\partial \varphi'} A_r(r,\varphi') = \frac{\partial}{\partial r}\left[rA_\varphi(r,\varphi')\right] \text{ and } \frac{\partial}{\partial r}[rB(r)] + \frac{\partial^2}{\partial r\partial \varphi'} A_r(r,\varphi') = \frac{\partial^2}{\partial r^2}\left[rA_\varphi(r,\varphi')\right]$$

and substituting this in (3.7) and (3.8) and in derivatives with respect to $\varphi$ we obtain

$$\frac{\partial \Psi}{\partial r} = -\frac{ie}{\hbar c}(A_r(r,\varphi) - A_r(r,0))\Psi + Ce^{i\frac{\lambda\varphi}{\hbar}} e^{\frac{ie}{\hbar c}\int_0^r r'dr'B(r')\varphi} e^{-\frac{ie}{\hbar c}\int_0^\varphi rd\varphi' A_\varphi(r,\varphi')} \frac{\partial f}{\partial r} \quad (3.9)$$

$$\frac{\partial^2 \Psi}{\partial r^2} = -\frac{ie}{\hbar c}\left[\frac{\partial}{\partial r} A_r(r,\varphi) - \frac{\partial}{\partial r} A_r(r,0)\right]\Psi$$

$$-\frac{ie}{\hbar c}\left[A_r(r,\varphi) - A_r(r,0)\right]\frac{\partial \Psi}{\partial r}$$

$$+ C\frac{ie}{\hbar c} rB(r)\varphi e^{i\frac{\lambda\varphi}{\hbar}} e^{\frac{ie}{\hbar c}\int_0^r r'dr'B(r')\varphi} e^{-\frac{ie}{\hbar c}\int_0^\varphi rd\varphi' A_\varphi(r,\varphi')} \frac{\partial f}{\partial r} \quad (3.10)$$

$$- C\frac{ie}{\hbar c}\int_0^\varphi d\varphi' \frac{\partial}{\partial r}\left[rA_\varphi(r,\varphi')\right]e^{i\frac{\lambda\varphi}{\hbar}} e^{\frac{ie}{\hbar c}\int_0^r r'dr'B(r')\varphi} e^{-\frac{ie}{\hbar c}\int_0^\varphi rd\varphi' A_\varphi(r,\varphi')} \frac{\partial f}{\partial r}$$

$$+ Ce^{i\frac{\lambda\varphi}{\hbar}} e^{\frac{ie}{\hbar c}\int_0^r r'dr'B(r')\varphi} e^{-\frac{ie}{\hbar c}\int_0^\varphi rd\varphi' A_\varphi(r,\varphi')} \frac{\partial^2 f}{\partial r^2}$$

$$\frac{\partial \Psi}{\partial \varphi} = \left[ i\frac{\lambda}{\hbar} + \frac{ie}{\hbar c}\int_0^r r'dr'B(r') - \frac{ie}{\hbar c}rA_\varphi(r,\varphi) \right]\Psi \quad \text{(3.11)}$$

$$\begin{aligned}\frac{\partial^2 \Psi}{\partial \varphi^2} &= -\frac{ie}{\hbar c}r\frac{\partial}{\partial \varphi}A_\varphi(r,\varphi)\Psi + \left[ i\frac{\lambda}{\hbar} + \frac{ie}{\hbar c}\int_0^r r'dr'B(r') - \frac{ie}{\hbar c}rA_\varphi(r,\varphi) \right]\frac{\partial \Psi}{\partial \varphi} \\ &= -\frac{ie}{\hbar c}r\frac{\partial}{\partial \varphi}A_\varphi(r,\varphi)\Psi + \left[ i\frac{\lambda}{\hbar} + \frac{ie}{\hbar c}\int_0^r r'dr'B(r') - \frac{ie}{\hbar c}rA_\varphi(r,\varphi) \right]^2 \Psi \end{aligned} \quad \text{(3.12)}$$

Finally, we substitute all the above derivatives in the Schrodinger eigenvalue equation to get

$$\begin{aligned}
&\frac{\hbar ie}{2mc}\left[\frac{\partial}{\partial r}A_r(r,\varphi) - \frac{\partial}{\partial r}A_r(r,0)\right]\Psi + \frac{\hbar ie}{2mc}\left[A_r(r,\varphi) - A_r(r,0)\right]\frac{\partial \Psi}{\partial r} \\
&-C\frac{\hbar ie}{2mc}rB(r)\varphi e^{i\frac{\lambda \varphi}{\hbar}} e^{\frac{ie}{\hbar c}\int_0^r r'dr'B(r')\varphi} e^{-\frac{ie}{\hbar c}\int_0^\varphi rd\varphi' A_\varphi(r,\varphi')} \frac{\partial f}{\partial r} + C\frac{\hbar ie}{2mc}\int_0^\varphi d\varphi' \frac{\partial}{\partial r}\left[rA_\varphi(r,\varphi')\right]e^{i\frac{\lambda \varphi}{\hbar}} e^{\frac{ie}{\hbar c}\int_0^r r'dr'B(r')\varphi} e^{-\frac{ie}{\hbar c}\int_0^\varphi rd\varphi' A_\varphi(r,\varphi')} \frac{\partial f}{\partial r} \\
&-\frac{\hbar^2}{2m}Ce^{i\frac{\lambda \varphi}{\hbar}}e^{\frac{ie}{\hbar c}\int_0^r r'dr'B(r')\varphi} e^{-\frac{ie}{\hbar c}\int_0^\varphi rd\varphi' A_\varphi(r,\varphi')} \frac{\partial^2 f}{\partial r^2} \\
&+\frac{\hbar ie}{2mrc}(A_r(r,\varphi) - A_r(r,0))\Psi - \frac{\hbar^2}{2mr}Ce^{i\frac{\lambda \varphi}{\hbar}}e^{\frac{ie}{\hbar c}\int_0^r r'dr'B(r')\varphi} e^{-\frac{ie}{\hbar c}\int_0^\varphi rd\varphi' A_\varphi(r,\varphi')} \frac{\partial f}{\partial r} \\
&+\frac{\hbar ie}{2mrc}\frac{\partial}{\partial \varphi}A_\varphi(r,\varphi)\Psi - \frac{\hbar^2}{2mr^2}\left[i\frac{\lambda}{\hbar} + \frac{ie}{\hbar c}\int_0^r r'dr'B(r') - \frac{ie}{\hbar c}rA_\varphi(r,\varphi)\right]^2 \Psi \\
&-\frac{e^2}{mc^2}A_r(r,\varphi)(A_r(r,\varphi) - A_r(r,0))\Psi - i\hbar\frac{e}{mc}Ce^{i\frac{\lambda \varphi}{\hbar}}e^{\frac{ie}{\hbar c}\int_0^r r'dr'B(r')\varphi} e^{-\frac{ie}{\hbar c}\int_0^\varphi rd\varphi' A_\varphi(r,\varphi')} \frac{\partial f}{\partial r} \\
&-i\hbar\frac{e}{mcr}A_\varphi(r,\varphi)\left[i\frac{\lambda}{\hbar} + \frac{ie}{\hbar c}\int_0^r r'dr'B(r') - \frac{ie}{\hbar c}rA_\varphi(r,\varphi)\right]\Psi - i\hbar\frac{e}{2mc}\vec{\nabla}\cdot\vec{A}\Psi + \frac{e^2}{2mc^2}A^2\Psi = E\Psi
\end{aligned}$$
(3.13)

And after a number of further algebraic manipulations, we obtain the following simplified result

$$\begin{aligned}
&\left(-\frac{\hbar ie}{2mc}\frac{\partial A_r(r,0)}{\partial r} + \frac{e^2}{2mc^2}A_r^2(r,0) - \frac{\hbar ie}{2mrc}A_r(r,0) + \frac{1}{2mr^2}\left(\lambda + \frac{e}{c}\int_0^r r'dr'B(r')\right)^2\right)f - \frac{\hbar ie}{mc}A_r(r,0)\frac{\partial f}{\partial r} \\
&-\frac{\hbar^2}{2m}\frac{\partial^2 f}{\partial r^2} - \frac{\hbar^2}{2mr}\frac{\partial f}{\partial r} = Ef
\end{aligned} \quad \text{(3.14)}$$

which, upon transforming $f(r)$ as $f(r) = e^{-\frac{ie}{\hbar c}\int_0^r dr' A_r(r',0)} X(r)$ leads to

$$-\frac{\hbar^2}{2mr}\frac{\partial}{\partial r}\left(r\frac{\partial X}{\partial r}\right) + \frac{1}{2mr^2}\left(\lambda + \frac{e}{c}\int_0^r r'dr'B(r')\right)^2 X = EX \text{ , (3.15)}$$

which is in turn a generalized Laguerre equation. The full wavefunction can then be written as:

$$\Psi(r,\varphi) = Ce^{i\frac{\lambda\varphi}{\hbar}} e^{\frac{ie}{\hbar c}\int_0^r r'dr'B(r')\varphi} e^{-\frac{ie}{\hbar c}\left[\int_0^\varphi rd\varphi' A_\varphi(r,\varphi') + \int_0^r dr' A_r(r',0)\right]} X(r) \text{ (3.16)}$$

*This is a pure quantum solution of the time-independent Schrodinger equation for a general radially-varying magnetic field.* Note that the appearing phase factor again constitutes of a magnetic flux plus a line integral (along a connected path) of the vector potential $\vec{A}$. If $\vec{B}(\vec{r}) = \vec{B}$, namely, when the magnetic field is homogeneous, eq. (3.15) becomes

$$-\frac{\hbar^2}{2mr}\frac{\partial}{\partial r}\left(r\frac{\partial X}{\partial r}\right) + \frac{1}{2mr^2}\left(\lambda + \frac{eBr^2}{2c}\right)^2 X = EX \text{ (3.17)}$$

which is the same with eq. (2.14) that we obtained in last section for a uniform $B$, with solution (eq. (3.16))

$$\Psi(r,\varphi) = Ce^{i\frac{\lambda\varphi}{\hbar}} e^{\frac{ieBr^2}{2\hbar c}\varphi} e^{-\frac{ie}{\hbar c}\left[\int_0^\varphi rd\varphi' A_\varphi(r,\varphi') + \int_0^r dr' A_r(r',0)\right]} X(r), \text{ (3.18)}$$

where $X(r)$ are Laguerre polynomials. In addition, note that when the vector potential has only azimuthal component, namely

$$A_\varphi = \frac{1}{r}\int_0^r dr' r' B(r') \text{ and } A_r = 0 \text{ (3.19)}$$

then the pseudo-angular momentum reads $L_z = -i\hbar\frac{\partial}{\partial\varphi} + \frac{e}{c}\int_0^r dr' r' B(r') - \frac{e}{c}\int_0^r B(r')r'dr' = -i\hbar\frac{\partial}{\partial\varphi}$, which is equal to the canonical angular momentum; the wavefunction (3.16) then becomes simply

$$\Psi(r,\varphi) = Ce^{i\frac{\lambda\varphi}{\hbar}} X(r) \text{ (3.20)}$$

which must be single-valued upon azimuthal trips by $2\pi$, and therefore the angular momentum remains quantized ($\lambda$ is an integer multiple of $\hbar$, as expected). As for the use of the second constant of motion, the generalized pseudomomentum

$$\vec{K} = \vec{\Pi} + \frac{e}{c}\int d\vec{r} \times \vec{B} \quad (3.21)$$

which is again path-dependent, it is convenient to solve the problem in the case of magnetic fields that depend only on a single Cartesian variable, for example $\vec{B} = \vec{B}(x)$ (a striped pattern, with vertical stripes), in which case we obtain

$$K_x = \Pi_x + \frac{e}{c}B(x)y \quad (3.22)$$

$$K_y = \Pi_y - \frac{e}{c}\int_0^x B(x')dx' \quad (3.23)$$

Direct algebraic manipulations give $[H, K_y] = 0$, while for the $x$ coordinate we get

$$[H, K_x] = -\frac{i\hbar e}{2mc}\left[\Pi_x y \frac{\partial B}{\partial x} + y \frac{\partial B}{\partial x}\Pi_x\right] \neq 0 \quad (3.24)$$

(i.e. the pseudomomentum along the direction perpendicular to the stripes is not conserved).
If the magnetic field were solely dependent on $y$ (hence now the stripes being horizontal), then we would have $[H, K_x] = 0$ and $[H, K_y] \neq 0$. Proceeding with this latter case as an example, we can therefore choose $H$ and $K_x$ and try to find their common eigenfunctions, which turn out to be the following

$$\Psi(x,y) = Ce^{i\left(\frac{k_x}{\hbar} - \frac{e}{\hbar c}\int_0^y B(y')dy'\right)x} e^{-\frac{ie}{\hbar c}\int_0^x dx' A_x(x',y)} f(y) = Ce^{i\frac{k_x x}{\hbar}} e^{-i\frac{ex}{\hbar c}\int_0^y B(y')dy'} e^{-\frac{ie}{\hbar c}\left[\int_0^x dx' A_x(x',y) + \int_0^y dy' A_y(0,y')\right]} Y(y)$$

(3.25)

with $Y$ satisfying

$$-\frac{\hbar^2}{2m}\frac{\partial^2 Y}{\partial y^2} + \frac{1}{2m}\left(k_x - \frac{e}{c}\int_0^y B(y')dy'\right)^2 Y = EY \quad (3.26)$$

and with $-\infty < k_x < \infty$ the pseudomomentum $K_x$-eigenvalues. And now note that the lower limit of the integrals appearing in these equations can arbitrarily be any constant number (an initial point of the path) that can be denoted by $(x_0, y_0)$, the final results being

$$\Psi(x,y) = Ce^{i\frac{k_x x}{\hbar}} e^{-i\frac{e(x-x_0)}{\hbar c}\int_{y_0}^y B(y')dy'} e^{-\frac{ie}{\hbar c}\left[\int_{x_0}^x dx' A_x(x',y) + \int_{y_0}^y dy' A_y(x_0,y')\right]} Y(y) = Ce^{i\frac{k_x x}{\hbar}} e^{-i\frac{e}{\hbar c}\Lambda(x,y)} Y(y)$$

(3.27)

with $\Lambda(x,y) = \int_{x_0}^{x}\int_{y_0}^{y} B(y')dx'dy' + \int_{x_0}^{x} dx' A_x(x',y) + \int_{y_0}^{y} dy' A_y(x_0,y')$, which in simply-connected space must be a single-valued phase function, and with $Y$ satisfying

$$-\frac{\hbar^2}{2m}\frac{\partial^2 Y}{\partial y^2} + \frac{1}{2m}\left(k_x - \frac{e}{c}\int_{y_0}^{y} B(y')dy'\right)^2 Y = EY.$$ Eq. (3.27) represents a solution (at any arbitrary gauge) of the static Schrodinger equation that has a well-defined $x$-component of the pseudomomentum. One notes that $\vec{\nabla}\Lambda$ is indeed our arbitrary $\vec{A}$ minus the Bawin-Burnel gauge [11], hence the above $\Lambda$ takes us from the Bawin-Burnel gauge (viewed as a "universal reference gauge") to any arbitrary gauge $\vec{A} = A_x \hat{i} + A_y \hat{j}$ that one wishes.

We should point out that considerations that generalize the ones of last section (regarding wavefunctions defined in different gauges) apply here as well. For example, if we choose the particular gauge $A_x = 0$ and $A_y = xB(y)$ the phase factor in eq. (3.27) gives a $\Lambda$ of the form

$$\Lambda(x,y) = \int_{x_0}^{x}\int_{y_0}^{y} B(y')dx'dy' + \int_{y_0}^{y} dy' x_0 B(y') = x\int_{y_0}^{y} dy' B(y') \quad \textbf{(3.28)}$$

and (3.27) reads $\Psi(x,y) = Ce^{i\frac{k_x x}{\hbar}} e^{-i\frac{e}{\hbar c} x \int_{y_0}^{y} dy' B(y')} Y(y)$, **(3.29)**

while if we choose another gauge $A_x = -\int_{y_0}^{y} dy' B(y')$ and $A_y = 0$ (3.27) gives

$$\Psi(x,y) = Ce^{i\frac{k_x x}{\hbar}} Y(y) \quad \textbf{(3.30)}$$

Note that, once again, (3.29) and (3.30) only differ by the correct (single) phase factor (that is straightforward to see that it indeed connects the two gauges, namely the difference of the two vector potentials is indeed the *grad* of the above $\Lambda$). We see therefore that if one is sufficiently careful, the cautionary remark of Swenson does not apply (or can be made inapplicable), *even in cases of inhomogeneous magnetic fields*.

Let us also briefly mention in passing a word on energy eigenvalues of the above eignefunctions. These can be generally determined through appropriate transfer matrix methods (see [12]), and it should be stressed that all previous workers in this field focus (quite justifiably) on determining these eigenvalues, claiming however (see i.e. the most recent and updated [12]) that the corresponding eigenvectors are not possible to determine in closed form – this giving an extra indication of the possible usefulness of the present work.

Finally, before ending the core of this article, we find it useful to include in a separate section a couple of comments that we find interesting for further study (related to the role of the point $(x_0,y_0)$), one on the probability flux vector (quantum mechanical current) in connection to the Hellmann-Feynman theorem [13], and one on Berry's phases [14] with respect to adiabatic and cyclic variations of the point $(x_0,y_0)$.

## *4. Probability current and Berry's phases*

Although the issues that will be discussed in this section are of independent interest and significance (i.e. they might be relevant to a much wider application, related to topologically nontrivial quantum systems) we will continue for concreteness using the final example of last section. First we recall that the local probability flux vector (or quantum probability current density) is generally defined as

$$\vec{J}_{loc} = \text{Re}\left(\Psi^* \frac{\vec{\Pi}}{m} \Psi\right) = \frac{i\hbar}{2m}\left[\Psi \vec{\nabla} \Psi^* - \Psi^* \vec{\nabla} \Psi\right] + \frac{e}{mc}\vec{A}|\Psi|^2 \quad (4.1)$$

that, for the particular case (of $B=B(y)$) and in the second gauge (with $A_y = 0$) of the previous section, has components

$$J_{locx}(y) = \frac{i\hbar}{2m}\left[\Psi \frac{\partial}{\partial x}\Psi^* - \Psi^* \frac{\partial}{\partial x}\Psi\right] + \frac{e}{mc}A_x|\Psi|^2 = \frac{1}{m}\left[k_x - \frac{e}{c}\int_{y_0}^{y} B(y')dy'\right]|Y(y)|^2 \quad (4.2)$$

$$J_{locy} = \frac{i\hbar}{2m}\left[\Psi \frac{\partial}{\partial y}\Psi^* - \Psi^* \frac{\partial}{\partial y}\Psi\right] + \frac{e}{mc}A_y|\Psi|^2 = \frac{i\hbar}{2m}\left[Y(y)\frac{\partial}{\partial y}Y^*(y) - Y^*(y)\frac{\partial}{\partial y}Y(y)\right] \quad (4.3)$$

Applying the well-known Hellmann-Feynman theorem [13] with respect to parameter $y_0$ in its ordinary form (note however on possible extra non-Hermitian boundary contributions which we leave out that might play a role, see [15]) we obtain

$$\frac{\partial H}{\partial y_0} = \frac{eB(y_0)}{mc}\left(k_x - \frac{e}{c}\int_{y_0}^{y} B(y')dy'\right) \quad (4.4)$$

and therefore

$$\frac{\partial E}{\partial y_0} = \left\langle \frac{\partial H}{\partial y_0}\right\rangle \Rightarrow \frac{\partial E}{\partial y_0} = \frac{eB(Y_0)}{c}\int_{-\infty}^{+\infty} dy J_{locx}(y) \quad (4.5)$$

We see that the eigenenergies depend explicitly on $y_0$ (which is actually not a surprise, although one would think that changing $y_0$ is equivalent to changing the vector potential by a constant, see ref. [16]) except when $B(y_0) = 0$ (i.e. the initial point is outside the magnetic field) or, for $B$ uniform, whenever $\int_{-\infty}^{+\infty} dy J_{locx}(y) = 0$ holds.

Using then the arbitrary gauge-wavefunction (3.27)

$$\Psi(x,y) = C e^{i\frac{k_x x}{\hbar}} e^{-i\frac{e}{\hbar c}\Lambda(x,y)} Y(y)$$

with $\Lambda(x,y) = \int_{x_0}^{x}\int_{y_0}^{y} B(y')dx'dy' + \int_{x_0}^{x} dx' A_x(x',y) + \int_{y_0}^{y} dy' A_y(x_0,y')$

we can interpret the initial point $(x_0, y_0)$ as a slowly varying parameter. A quantity then that may acquire a special significance is the Berry's phase picked up by the particle's wavefunction during cyclic adiabatic changes of $(x_0, y_0)$; in our case this can be determined analytically, and it is equal to

$$\gamma = i \oint d\vec{R}.\langle\Psi|\vec{\nabla}_{\vec{R}}|\Psi\rangle, \text{ with } \vec{R} = (x_0, y_0) \quad \textbf{(4.6)}$$

Performing the necessary calculations we have

$$\frac{\partial \Psi}{\partial y_0} = -i\frac{e}{\hbar c}\frac{\partial \Lambda}{\partial y_0}\Psi + C e^{i\frac{k_x x}{\hbar}} e^{-i\frac{e}{\hbar c}\Lambda(x,y)} \frac{\partial Y}{\partial y_0} \quad \textbf{(4.7)}$$

and $\frac{\partial \Psi}{\partial x_0} = -i\frac{e}{\hbar c}\frac{\partial \Lambda}{\partial x_0}\Psi$, **(4.8)**

with $\frac{\partial \Lambda}{\partial y_0} = -B(y_0)(x - x_0) - A_y(x_0, y_0)$ και $\frac{\partial \Lambda}{\partial x_0} = -A_x(x_0, y_0)$ **(4.9)**

$$\frac{\partial \Psi}{\partial y_0} = i\frac{e}{\hbar c}\left(B(y_0)(x-x_0) + A_y(x_0,y_0)\right)\Psi + C e^{i\frac{k_x x}{\hbar}} e^{-i\frac{e}{\hbar c}\Lambda(x,y)} \frac{\partial Y}{\partial y_0} \quad \textbf{(4.10)}$$

$$\frac{\partial \Psi}{\partial x_0} = i\frac{e}{\hbar c} A_x(x_0, y_0)\Psi \quad \textbf{(4.11)}$$

Choosing again $B(y_0) = 0$, we have $\partial Y/\partial y_0 = 0$ and the Berry's phase can be written as

$$\gamma = i\oint d\vec{R}.\langle\Psi|\vec{\nabla}_{\vec{R}}|\Psi\rangle = -\frac{e}{\hbar c}\oint d\vec{R}.\vec{A}(x_0,y_0) + i\oint d\vec{R}.\left\langle Y\left|\frac{\partial Y}{\partial y_0}\right.\right\rangle \hat{j}. \quad \textbf{(4.12)}$$

Now, because $\left\langle Y\left|\frac{\partial Y}{\partial y_0}\right.\right\rangle = -\left\langle \frac{\partial Y}{\partial y_0}\left|Y\right.\right\rangle$ we may write (4.12) as

$$\gamma = i\oint d\vec{R}.\langle\Psi|\vec{\nabla}_{\vec{R}}|\Psi\rangle = -\frac{e}{\hbar c}\oint d\vec{R}.\vec{A}(x_0,y_0) + \frac{i}{2}\oint d\vec{R}\left[\left\langle Y\bigg|\frac{\partial Y}{\partial y_0}\right\rangle - \left\langle\frac{\partial Y}{\partial y_0}\bigg|Y\right\rangle\right]\hat{j}$$

$$= -\frac{e}{\hbar c}\oint d\vec{R}.\left[\vec{A}(x_0,y_0) + \frac{imc}{2e}\int_{-\infty}^{+\infty}dy J_{locy}\hat{j}\right]$$
, (4.13)

where the last step is rigorously valid if we further assume that $B(y')$ has a constant value for $y_0 < y' \leq y$ so that $Y$ is then a function of $y - y_0$ only, and therefore $\partial Y/\partial y = -\partial Y/\partial y_0$. The result (4.13) for the Berry's phase is equal to the Aharonov-Bohm phase plus a contribution from the global electric current picked up in the parameter-based cyclic loop (and it seems to generalize previous 1D results [17]).

The above is valid for the flat 2D problem. [Incidentally, we note here that, for the rotationally symmetric system $B=B(r)$, the result turns out to be (by conformal mapping) similar, namely an Aharonov-Bohm phase plus a global current (which for a closed system is zero) and it is essentially a generalization of the result for the well-known Berry's rigid box that is adiabatically moved around a flux [BERRY].] It should be noted, however, that if the above system has periodic boundary conditions (along a Cartesian direction) and it can equivalently be folded into a cylinder, then the Berry's phase picked up by trips around the empty space is a more interesting story: apart from the Aharonov-Bohm phase, the full Berry's phase also contains a term containing the global electric current [17], and this is a matter of further interest (the extra global electric current term can now differentiate (in this 2D problem – compared to the previously published 1D problems in refs [17]) between topologically non trivial (folded) and trivial (flat) cases (i.e. in the two cases, the extra global current term is of a different type, belonging to different so-called homotopy classes)), these being matters that need to be presented in more detail elsewhere.

## 5. *Inclusion of a homogeneous electric field*

Before ending this article, we digress a bit from the main systems considered, in order to address the question whether one could even incorporate into our pseudomomentum-based method the possible additional presence of an electric field. We will briefly see that the answer is affirmative, and that now we will have to follow extended procedures that will now involve the *t*-dependent (rather than static) Schrodinger equation. Indeed, let us focus on the special case where an additional homogeneous in-plane electric field $\vec{E}$ is imposed on the system. The Hamiltonian is now $H^E = H^{E=0} + e\,\vec{E}.\vec{r}$, and in this case, the generalized pseudomomentum (*cf.* eq. **(2.1)**) is extended (by consideration again of the Newton's equation of motion with the additional electric force -*e E*) to

$$\vec{K} = \vec{p} + \frac{e}{c}\vec{A} + \frac{e}{c}\vec{r}\times\vec{B} + e\vec{E}t \quad \textbf{(5.1)}$$

and it is no longer simply a constant of the motion as before (it does not commute with *H*) but it becomes an invariant operator [18], namely it obeys the following relation

$$\frac{\partial \vec{K}}{\partial t} = -\frac{i}{\hbar}\left[H, \vec{K}\right] \quad (5.2)$$

with $\frac{\partial \vec{K}}{\partial t} = e\vec{E}$. In this case, the *y*-component of $\vec{K}$ and *H* no longer share the same eigenfunctions (neither does the *x*-component of $\vec{K}$ and *H*), but the two eigenfunctions differ by a time-dependent phase factor. If we choose the *E*-field to point to the *x*-direction, we can use the eigenfunction of $K_y$ to find the eigenfunction of the Hamiltonian, which after some algebraic manipulations (extensions of the ones made previously) turns out to be of the form

$$\Psi(x,y) = e^{i\frac{k_y y}{\hbar}} e^{-i\frac{e}{\hbar c}\left[-xyB + \int_0^x dx' A_x(x',0) + \int_0^y dy' A_y(x,y') - c\int_0^t dt' V(x=0,y=0,t')\right]} g(x,t) \quad (5.3)$$

with *V* a time-dependent scalar potential defined at the origin, with $k_y$ an eigenvalue of $K_y$ and with the function $g(x,t)$ obeying the following differential equation

$$-\frac{\hbar^2}{2m}\frac{\partial^2 g}{\partial x^2} + \left[\frac{1}{2m}\left(k_y + \frac{eBx}{c}\right)^2 + exE_x\right]g = i\hbar\frac{dg}{dt} \quad (5.4)$$

Factoring then terms that include the *x*-variable results in a harmonic oscillator-type of time-dependent Schrodinger equation, namely

$$-\frac{\hbar^2}{2m}\frac{\partial^2 g}{\partial x^2} + \left[\frac{e^2 B^2}{2mc^2}(x-X_0)^2 + \frac{ck_y E_x}{B} - \frac{mc^2 E_x^2}{2B^2}\right]g = i\hbar\frac{dg}{dt} \quad (5.5)$$

with $X_0 = \frac{ck_y}{eB} - \frac{mc^2 E_x}{eB^2}$ the new guiding center operator, that is now drifting inside the crossed electric and magnetic fields with the well-known drift velocity $\vec{V}_D = c\frac{\vec{E}\times\vec{B}}{B^2}$. Thus, we have shown that no matter what the choice of the vector potential may be, we always end up with harmonic oscillator wavefunctions and eigenenergies, namely

$$g(x,t) \sim H_n\left(\frac{x-X_o}{l_B}\right) e^{-\frac{(x-X_o)^2}{2l_B^2} - \frac{i\varepsilon_{n,k_y} t}{\hbar}} \quad (5.6) \text{ and}$$

$$\varepsilon_{n,k_y} = \hbar\omega(n+1/2) - \frac{mc^2 E_x^2}{2B^2} + \frac{ck_y}{B}E_x, \quad \text{with } \omega = \frac{eB}{mc}. \quad (5.7)$$

Equation (5.4) cures in a rather elegant way the "disparity" existing between the two Landau gauges when a homogeneous electric field is present in the system. For example, if the electric field is along the *x*-direction, and the magnetic field (always being along the *z*-direction) is described by the Landau gauge $\vec{A} = (0, Bx)$, the Hamiltonian is

$$H = \frac{1}{2m}\left(\vec{p} - eBx\hat{j}/c\right)^2 + exE_x \quad (5.8)$$

In this case, one can immediately write down the Hamiltonian in a more compact form (after factoring of the *x*-terms), namely

$$H = \frac{p_x^2}{2m} + \frac{e^2 B^2}{2mc^2}\left[x + \frac{mc^2}{e^2 B^2}\left(eE_x - \frac{p_y eB}{mc}\right)\right]^2 - \frac{mc^2 E_x^2}{2B^2} + \frac{cp_y}{B}E_x \quad (5.9)$$

and, by observing that $p_y$ is a constant of motion, the eigenenergies are

$$\varepsilon = \hbar\omega(n+1/2) - \frac{mc^2 E_x^2}{2B^2} + \frac{cp_y}{B}E_x \quad (5.10)$$

(as in (5.7)), while the wavefunctions in the *x*-direction are as in (5.6) with the Hermite polynomials centered at $X_0 = -mc^2 E_x / eB^2 + cp_y / eB$. This is all well-known. The "disparity" arises when one tries to solve the problem using the other Landau gauge, $\vec{A} = (-By, 0)$ with the resulting Hamiltonian being

$$H = \frac{1}{2m}\left(\vec{p} + eBy\hat{i}/c\right)^2 + exE_x \quad (5.11)$$

In this case, there are two terms that depend on different spatial variables and as a result they cannot be factored into a perfect square. The problem looks rather awkward to handle and people always seem to select the former case with only one spatial variable. However, we have given here equations (5.3), (5.5), (5.6) and (5.7) and these are the general solutions *for every vector potential choice*, hence covering also the above awkward choice of gauge. Equation (5.1) can also be extended to the more general cases of spatial inhomogeneities of both the magnetic and electric fields, and even for *t*-dependent electric field (in this case the product of the last term will become a time-integral, i.e. we will now have $\vec{K} = \vec{p} + \frac{e}{c}\vec{A}_{\vec{B}} + \frac{e}{c}\int_C d\vec{r} \times \vec{B}(\vec{r}) + e\int_0^t \vec{E}(t')dt'$), and in all cases one can obtain the general form of the wavefunctions (with similar procedures of simultaneous diagonalizations with *H* as earlier). In all cases, it seems that the extra phases that show up at the end results have the general form of the non-local terms of refs [5,6] involving also time-integrals of scalar potentials, surface integrals of the magnetic field, and mixed temporal and spatial integrals of the electric field (this essentially being a generalization of the electric Aharonov-Bohm effect). All this needs, however, a more detailed and self-contained presentation which we leave for the future.

## *6. Conclusions*

We have shown that when working with Landau gauges (or, actually any other gauge), the wavefunction that corresponds to one gauge can be easily transformed to a wavefunction that corresponds to another gauge by a simple (single-phase) gauge transformation, as long as the *same* Cartesian component of pseudomomentum (or pseudo-angular momentum) is simultaneously diagonalized with $H$ in **each** *of the two gauges*. This may sound natural, but it is *not* what is being followed in textbooks or in the research literature (where *canonical* momenta are used as constants of the motion instead of pseudomomentum). The handicap of the standard procedure is that only one Cartesian component of the canonical momentum happens to be a constant of the motion *only* in one Landau gauge, and only the other Cartesian component of the canonical momentum happens to be a constant of the motion only in the other Landau gauge, not both components. In contradistinction, both components of the pseudomomentum are always constants of the motion, in *any* gauge, as shown clearly in the text, hence the pseudomomentum providing a more powerful book-keeping that eliminates the complications. If one chooses to solve the problem without paying attention to the pseudomomentum, it may happen that upon change of gauge there will be different components of $\vec{K}$ diagonalized in the different gauges (as it happens implicitly in the standard procedures that mimic the original solution of Landau with use of the canonical momenta), and then the two wavefunctions will not be connected by a single-phase relation, but through a linear combination on all quantum numbers that lie within a single degeneracy-subspace (i.e. correspond to the same energy) but are labeled by only a "partial" constant of motion – this is what occurred in Swenson's warning article [1].

Most importantly, however, we have also noted that similar conclusions can also be drawn even in the more difficult cases of inhomogeneous magnetic fields (an area that has so far focused on the energy eigenvalues – with a noted difficulty in getting information on the corresponding eigenfunctions), and in such cases not only did we provide closed analytical forms for the wavefunctions, but we additionally calculated Berry's phases (for the flat 2D Landau problem) upon small variations of the location of an initial point (that could be taken as the origin of coordinates). [This could be viewed as an innocent gauge transformation; note, however, that if we had periodic boundary conditions and had folded the problem along one direction, this change of initial point would no longer be so innocent – it would correspond to a singular gauge transformation with subtle physical consequences (a hidden transfer of physical momenta between the two systems), the simplest one being a gauge proximity effect [16].] In this case, because both $\vec{K}$ and $\vec{L}_z$ are expressed as line integrals and therefore depend on the path connecting the initial and final points, extra care must be taken in the sense that the integrand quantities must be curl free. Independence from the path of integration then means that these quantities are well defined constants of motion. We demonstrated in detail how utilization of such constants of motion can quickly lead to general solutions of the Schrodinger equation in certain nonuniform-field cases, making apparent how a careful use of gauge transformation techniques (combined with the proper use of pseudomomentum quantities) can be a powerful tool for the quick solution of difficult problems of this type. Finally, an inclusion of an in-plane electric field (even a *t*-dependent one) was briefly considered, with an even further generalized pseudomomentum defined in the text, demonstrating that our method is generalizable to treat even broader physical systems.